\documentclass{aa}
\usepackage{psfig}

\begin{document}

\title{Unveiling the nature of {\it INTEGRAL} objects through optical 
spectroscopy. IV. A study of six new hard X--ray sources\thanks{Based 
on observations collected 
at the Astronomical Observatory of Bologna in Loiano (Italy)
and with the 2.2m ESO/MPG telescope of the European Southern
Observatory in La Silla (Chile).}}

\author{N. Masetti\inst{1},
L. Bassani\inst{1},
A. Bazzano\inst{2}, 
A.J. Bird\inst{3},
A.J. Dean\inst{3},
A. Malizia\inst{1},
L. Norci\inst{4},
E. Palazzi\inst{1},
A.D. Schwope\inst{5},
J.B. Stephen\inst{1},
P. Ubertini\inst{2} and
R. Walter\inst{6}
}

\institute{
INAF -- Istituto di Astrofisica Spaziale e Fisica Cosmica di 
Bologna, Via Gobetti 101, I-40129 Bologna, Italy (formerly IASF/CNR,
Bologna)
\and
INAF -- Istituto di Astrofisica Spaziale e Fisica Cosmica di
Roma, Via Fosso del Cavaliere 100, I-00133 Rome, Italy (formerly 
IASF/CNR, Rome)
\and
School of Physics \& Astronomy, University of Southampton, Southampton, 
Hampshire, SO17 1BJ, United Kingdom  
\and
School of Physical Sciences, Dublin City University, Glasnevin, Dublin 
9, Republic of Ireland
\and
Astrophysikalisches Institut Potsdam, An der Sternwarte 16, D-14482 
Potsdam, Germany 
\and
INTEGRAL Science Data Centre, Chemin d'Ecogia 16, CH-1290 Versoix,
Switzerland
}

\offprints{N. Masetti (\texttt{masetti@iasfbo.inaf.it)}}
\date{Received 28 February 2006; accepted 21 April 2006}

\abstract{
We present further results from our onging optical spectrophotometric
campaign at the Astronomical Observatory of Bologna in Loiano (Italy) 
on unidentified hard X--ray sources detected by {\it INTEGRAL}. We
observed spectroscopically the putative optical counterparts of
the {\it INTEGRAL} sources IGR J00234+6141, IGR J01583+6713, IGR 
J06074+2205, IGR J13091+1137 and IGR J20286+2544. We find that the first 
two are Galactic objects, namely a 
Cataclysmic Variable at a distance $d \sim$ 300 pc and a Be/X transient 
High-Mass X--ray Binary (HMXB) located at $\sim$6.4 kpc, respectively,
whereas the last one is identified with MCG +04-48-002, a Starburst/H 
{\sc ii} galaxy at redshift $z$ = 0.013 hiding a Seyfert 2 
nucleus. We identify IGR J13091+1137 as the (likely Seyfert 2 type) 
active nucleus of galaxy NGC 4992, which we classify as an X--ray 
Bright, Optically Normal Galaxy; this is the first example of this 
type of object to be 
detected by {\it INTEGRAL}, and one of the closest of this class.
We moreover confirm the possible Be/X nature of IGR J06074+2205,
and we estimate it to be at a distance of $\sim$1 kpc.
We also reexamine the spectrum of the $z$ = 0.087 elliptical 
radio galaxy PKS 0352$-$686, the possible counterpart of the 
{\it INTEGRAL} 
source IGR J03532$-$6829, and we find that it is a BL Lac.
Physical parameters for these sources are also evaluated by 
discussing our findings in the context of the available multiwavelength 
information. These identifications further stress the importance of 
{\it INTEGRAL} in the study of the hard X--ray spectrum of Active 
Galactic Nuclei, HMXBs and Cataclysmic Variables.

}

\titlerunning{The nature of six hard X--ray {\it INTEGRAL} sources}
\authorrunning{N. Masetti et al.}

\maketitle

\section{Introduction}

Since its launch in October 2002, the {\it INTEGRAL} satellite (Winkler 
et al. 2003) is revolutionizing our knowledge of the hard X--ray
sky above 20 keV in terms of both sensitivity and positional accuracy
of the detected hard X--ray sources. Indeed, thanks to the capabilities 
of the ISGRI detector of the instrument IBIS (Ubertini et al. 2003), 
{\it INTEGRAL} is able to detect hard X--ray objects at the mCrab level 
with a typical localization accuracy of 2--3$'$ (Gros et al. 2003).
This has made it possible, for the first time, to obtain all-sky maps
in the 20--100 keV range with arcminute accuracy and down to mCrab 
sensitivities (e.g., Bird et al. 2004, 2006).

Most of the sources in these surveys are known Galactic X--ray binaries 
($\sim$50\%); however, a non-negligible fraction consists of background 
Active Galactic Nuclei (AGNs; $\sim$10\%) and Cataclysmic Variables 
(CVs; $\sim$5\%). A large part of the remaining objects (about one 
quarter of the sample) has no obvious
counterpart at other wavelengths and therefore cannot immediately
be associated with any known class of high-energy emitting objects.

However, it should be noted that, although circumstantial evidence 
was found (Dean et al. 2005) that most of these unknown objects are 
likely High-Mass X--ray Binaries (HMXBs), optical spectroscopy 
demonstrates that about half of them are AGNs (Masetti et al. 2004, 
2006a,b; hereafter Papers I, II and III).
Besides, despite the fact that substantial hard X--ray emission from CVs 
is not observed in general (e.g., de Martino et al. 2004 and references 
therein), {\it INTEGRAL} has already proved its capability to detect
this kind of source (e.g., Masetti et al. 2005, 2006c).

In our continuing effort to identify unknown {\it INTEGRAL} sources (see 
Papers I-III), we selected a further sample of five objects for which 
bright candidates could be pinpointed on the basis of their association 
with sources at other wavebands, mainly in soft X--rays and radio.
In one more case we considered, as a putative counterpart, a bright 
emission-line star within the ISGRI error circle. For the caveats 
regarding this latter choice, we refer the reader to Paper III.

We present here the spectroscopic results obtained at the 1.5-metre `G.D. 
Cassini' telescope of the Astronomical Observatory of Bologna on five of 
the selected sources, and we reexamine the optical spectroscopic 
data of the radio galaxy PKS 0352$-$686 acquired at ESO-La Silla (Chile) 
with the 2.2m ESO/MPG telescope. In Sect. 2 we introduce the sample of 
objects chosen for this observational campaign, while in Sect. 3 a 
description of the observations is given; Sect. 4 reports and discusses
the results for each source. Conclusions are drawn in Sect. 5.

In the following, when not explicitly stated otherwise, we will assume 
a Crab-like spectrum for our X--ray flux estimates, whereas for 
the {\it INTEGRAL} error boxes a conservative 90\% confidence level will 
be considered. When not mentioned, an {\it INTEGRAL} error box radius of 
3$'$ is assumed.

\section{The selected sources}

{\it IGR J00234+6141}. This weak hard X--ray source was detected by ISGRI
during a 1.6-Ms exposure centered on the Cassiopeia region of the Galaxy; 
the source flux is 0.72$\pm$0.12 mCrab in the 20--50 keV energy band, and
1.4$\pm$0.3 mCrab in the 50--100 keV band (den Hartog et al. 2006).
These authors suggest that, given that the source lies on the 
Galactic Plane ($b$ = $-$1$\fdg$0) and lacks a radio counterpart, 
it may be an X--ray binary.

The {\it INTEGRAL} source position reported by den Hartog et al. (2006) 
is RA = 00$^{\rm h}$ 23$^{\rm m}$ 24$^{\rm s}$ and Dec = +61$^{\circ}$ 
41$'$ 32$''$ (J2000). 
At the border of the {\it INTEGRAL} error box, and marginally consistent 
with it, is the bright {\it ROSAT} X--ray source 1RXS J002258.3+614111 
(Voges et al. 1999) with a 0.1--2.4 keV flux of 
(4.7$\pm$0.9)$\times$10$^{-13}$ erg cm$^{-2}$ s$^{-1}$. According to the 
findings of Stephen et al. (2005, 2006), this indicates that the source 
is the soft X--ray counterpart of IGR J00234+6141. Within the 11$''$ {\it 
ROSAT} error box (Fig. 1, upper left panel), at least 6 objects can be 
seen on the DSS-II-Red 
Survey\footnote{available at \texttt{http://archive.eso.org/dss/dss/}}; 
the two brightest optical sources have magnitudes $B \sim$ 16.6 and 
$R\sim$ 16.3 (the westernmost one), and $B \sim$ 17.2 and $R\sim$ 16.5 
(the easternmost one) according to the USNO-A2.0\footnote{available at \\ 
{\tt http://archive.eso.org/skycat/servers/usnoa}} catalogue.
We acquired spectra of each of these two objects to see whether either 
may be responsible for the hard X--ray emission detected by 
{\it INTEGRAL}.

Preliminary spectroscopic results on the possible counterpart of this 
source were already reported by Halpern \& Mirabal (2006) and Bikmaev 
et al. (2006), and are based on more recent spectroscopic observations 
compared to those shown in the present paper.

{\it IGR J01583+6713}. This new hard X--ray transient was discovered 
(Steiner et al. 2005) in early December 2005 with ISGRI, at coordinates 
(J2000) RA = 01$^{\rm h}$ 58$^{\rm m}$ 17$^{\rm s}$, Dec = 
+67$^{\circ}$ 13$'$ 12$''$, with an uncertainty of 2$'$. The flux 
at the time of the discovery was $\sim$14 mCrab in the 20--40 keV band, 
declining in the days following the discovery. The source was not 
detected down to a flux limit of 7 mCrab in the 40--80 keV band.

Pointed follow-up {\it Swift}/XRT observations of the field allowed the 
detection of the soft X--ray counterpart of IGR J01583+6713, at 
RA = 01$^{\rm h}$ 58$^{\rm m}$ 18$\fs$2, Dec =+67$^{\circ}$ 13$'$ 
25$\farcs$9 (J2000; Kennea et al. 2005). This localization with 
arcsecond precision allowed 
these authors to propose a single, relatively bright optical object lying 
within the {\it Swift}/XRT error box as the counterpart to this 
{\it INTEGRAL} source. This object (Fig. 1, upper middle panel) has 
magnitudes $B$ = 14.98, $R$ = 13.25, $I$ = 12.12 (Monet et al. 2003). 

Spectral fitting of {\it Swift}/XRT data of this source (Kennea et al. 
2005) indicates a 0.2--10 keV flux of 1.5$\times$10$^{-11}$ erg cm$^{-2}$ 
s$^{-1}$ and suggests the presence of a large neutral hydrogen column 
density, as high as 10$^{23}$ cm$^{-2}$, larger than the Galactic value 
in the direction of IGR J01583+6713.

The optical spectroscopic follow-up of Halpern \& Tyagi (2005a) confirmed
that this object is the counterpart of IGR J01583+6713 and classified 
the source as a Be/X--ray binary. In this case also, the identification 
was obtained through more recent spectroscopy compared to that 
presented here.

{\it IGR J03532$-$6829}. This object was detected in the {\it INTEGRAL} 
Large Magellanic Cloud survey of G\"otz et al. (2006), with coordinates 
(J2000) RA = 03$^{\rm h}$ 53$^{\rm m}$ 14$^{\rm s}$ and Dec = 
$-$68$^{\circ}$ 29$'$ 00$''$ (J2000) and a 20--40 keV flux of 0.6 mCrab. 
A {\it ROSAT} bright soft X--ray source (1RXS J035257.7$-$683120; Voges et 
al. 1999) and a SUMSS radio object (SUMSS J035257$-$683118; Mauch et al. 
2003) were found within the 3$\farcm$5-radius ISGRI error box.
Their positional errors were consistent with them being the same source. 
This object has an X--ray flux of (3.2$\pm$0.5)$\times$10$^{-12}$ erg 
cm$^{-2}$ s$^{-1}$ in the 0.1--2.4 keV band and a radio flux density 
of 390.8$\pm$11.8 mJy at 843 MHz. X--ray emission consistent with this 
position was also detected by {\it Einstein} (Elvis et al. 1992) and by 
{\it RXTE} (Revnivtsev et al. 2004), with fluxes 
(9.0$\pm$2.2)$\times$10$^{-12}$ erg cm$^{-2}$ s$^{-1}$ and 
(6.8$\pm$1.3)$\times$10$^{-12}$ erg cm$^{-2}$ s$^{-1}$ in the 0.16--3.5 
keV and 3--8 keV bands, respectively.

The radio and soft X--ray arcsecond-sized error boxes are 
coincident with the body of an elliptical galaxy (see Fig.1, upper 
right panel) characterized by the flat-spectrum radio source PKS 
0352$-$686 (Wright et al. 1994). The galaxy has USNO-A2.0 optical 
magnitudes $B \sim$ 
13.6 and $R \sim$ 12.5. According to the optical survey of Fischer et al. 
(1998), this galaxy has redshift $z$ = 0.087 and may be part of a cluster 
responsible for the X--ray emission, even if a BL Lac nature 
cannot be excluded. We thus performed a further analysis of the 
data reported by Fischer et al. (1998) on this object to try to ascertain 
its actual nature.

{\it IGR J06074+2205}. This transient source was discovered by {\it 
INTEGRAL}'s JEM-X telescopes and ISGRI instrument at coordinates (J2000) 
RA = 06$^{\rm h}$ 07$\fm$4, Dec = +22$^{\circ}$ 05$'$ (error radius: 
2$'$) between 15 and 16 February 2003 during an observation of the 
Crab region (Chenevez et al. 2004). According to these authors, the source 
had fluxes of 7$\pm$2 mCrab (3--10 keV) and $\sim$15 mCrab (10--20 keV) at 
the time of the discovery.

Recent (27 December 2005) optical spectroscopic observations of stars in 
the source error box show the presence of a Be star (Fig. 1, lower left 
panel) at RA = 06$^{\rm h}$ 07$^{\rm m}$ 26$\fs$6, Dec = +22$^{\circ}$ 
05$'$ 48$''$ with H$_\alpha$ emission (Halpern \& Tyagi 2005b).
These authors suggest that this is the counterpart of IGR J06074+2205 
and that the radio source reported by Pooley (2004) within the 
{\it INTEGRAL} error circle is an unrelated object.
We made detailed spectroscopic observations of the Be star to deepen 
the optical study of this object.

{\it IGR J13091+1137}. This is one of the 8 objects included in the
{\it INTEGRAL}/{\it Chandra} minisurvey of Sazonov et al. (2005),
with ISGRI coordinates RA = 13$^{\rm h}$ 09$^{\rm m}$ 04$\fs$1, Dec =
+11$^{\circ}$ 37$'$ 19$''$ (J2000). These authors report that this source
has 0.5--8 keV and 17--60 keV fluxes of (1.2$\pm$0.2)$\times$10$^{-12}$ 
erg cm$^{-2}$ s$^{-1}$ and (3.4$\pm$0.5)$\times$10$^{-11}$ erg cm$^{-2}$
s$^{-1}$, respectively, and a large neutral hydrogen column density, 
$N_{\rm H}$ = (90$\pm$10)$\times$10$^{22}$ cm$^{-2}$, along the line
of sight. 

The subarcsecond {\it Chandra} position, RA = 13$^{\rm h}$
09$^{\rm m}$ 05$\fs$60, Dec = +11$^{\circ}$ 38$'$ 02$\farcs$9 (J2000;
error radius: $\sim$0$\farcs$6) falls on the nucleus of the bright 
Sa-type spiral galaxy NGC 4992 (in Fig. 1, lower middle panel; see also 
Halpern 2005 
and Sazonov et al. 2005). This galaxy has total magnitude $B \sim$ 14.5
and lies at redshift $z$ = 0.0251$\pm$0.0002 (Prugniel 2006).
The nucleus of this galaxy is also a radio source, FIRST J130905.5+113803,
with 1.4 GHz flux density of 2.01$\pm$0.15 mJy (Becker et al. 1997).
Sazonov et al. (2005) thus suggested that this object is most likely an 
AGN, but no clearer indication on the actual nature of this source is 
reported by these authors. 

Preliminary results on our spectroscopic optical study of this source 
were given in Masetti et al. (2006d).

{\it IGR J20286+2544}. This source appears in the {\it INTEGRAL} AGN 
catalogue of Bassani et al. (2006), at coordinates (J2000) RA = 
20$^{\rm h}$ 28$^{\rm m}$ 37$\fs$4 and Dec = +25$^{\circ}$
45$'$ 54$''$ (J2000) with a 20--100 keV flux of 2.42$\pm$0.42 
mCrab. One single radio source, with flux density 27.4$\pm$1.2 mJy
at 1.4 GHz (Condon et al. 1998), is present in the X--ray error box. 
Coincident with this radio source, a relatively bright ($B \sim$ 15.4 mag) 
and close ($z$ = 0.0142$\pm$0.0002) galaxy, MCG +04-48-002, is found 
(Paturel et al. 2003; see Fig. 1, lower right panel). This object is also 
a bright IRAS far-infrared (FIR) galaxy (Sanders et al. 2003).

It should however be noted that, just outside the {\it INTEGRAL}
error circle, a further NVSS source at a 1.4 GHz flux density of
84.8$\pm$3.1 mJy is detected (Condon et al. 1998) positionally
coincident with the $B \sim$ 14.4 mag galaxy NGC 6921, lying at 
redshift $z$ = 0.0147$\pm$0.0007 (Paturel et al. 2003).

Neither of these galaxies was detected by {\it ROSAT} in the 0.1--2.4 keV 
band. However, 
an analysis of archival {\it Swift}/XRT data (Obs. ID: 00035276001; Landi 
et al., in preparation) shows that both objects are X--ray emitting in 
the 0.1--10 keV band. The arcsec-sized XRT error boxes clearly mark the 
nuclei of these galaxies as the sites of the detected X--ray emissions, 
with the nucleus of MCG +04-48-002 being about 10 times brighter than 
that of 
NGC 6921 in this X--ray range. Thus, given the position inside the {\it 
INTEGRAL} error circle and the X--ray intensity, MCG +04-48-002 is the 
best optical counterpart candidate of IGR J20286+2544. Nevertheless, 
optical spectroscopy is the optimal medium to tell which of the two 
galaxies is the responsible for the hard X--ray emission.

\section{Optical spectroscopy}

\begin{figure*}[t!]
\vspace{-5cm}
\hspace{-.5cm}
\centering{\mbox{\psfig{file=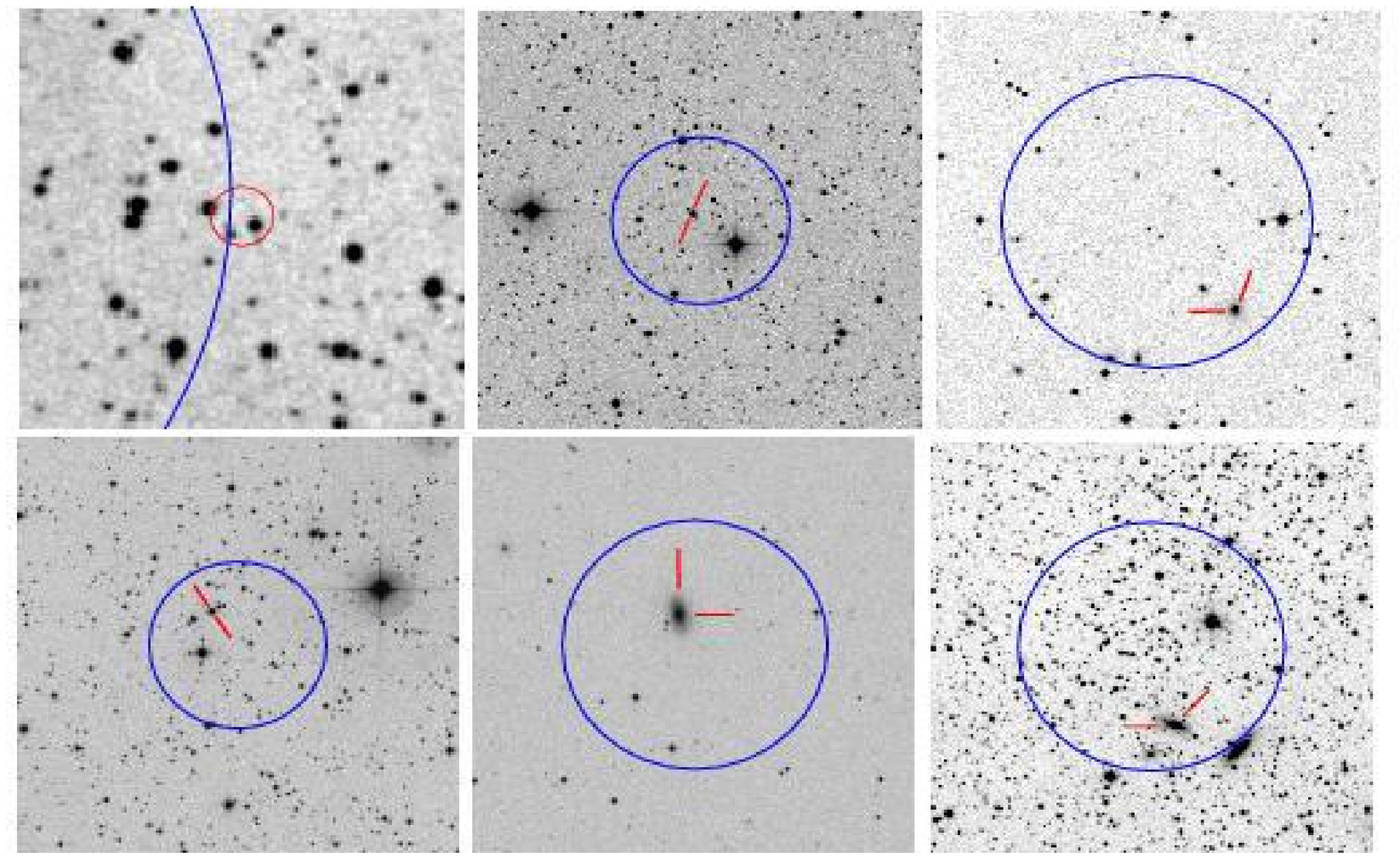,width=18cm}}}
\vspace{-9cm}
\caption{DSS-II-Red optical images of the fields of IGR J00234+6141 
(upper left panel), IGR J01583+6713 (upper middle panel),
IGR J03532$-$6829 (upper right panel), IGR J06074+2205 (lower left panel),
IGR J13091+1137 (lower middle panel) and IGR J20286+2544 
(lower right panel). In the upper left panel, 
the small circle indicates the {\it ROSAT} error box of the soft 
X--ray source 1RXS J002258.3+614111, whereas the larger curve
indicates a portion of the {\it INTEGRAL} error circle.
In the other panels, within the {\it INTEGRAL} error circles, 
the putative optical counterparts of IGR J01583+6713 and IGR J06074+2205,
as well as the galaxies PKS 0352$-$686, NGC 4992 and MCG +04-48-002, 
putative optical counterparts of IGR J03532$-$6829, IGR J13091+1137 and 
IGR J20286+2544, respectively, are indicated with tick marks.
In the lower right panel, southwest of the galaxy MCG +04-48-002, just 
outside the {\it INTEGRAL} error box, the galaxy NGC 6921 is present. 
Field sizes are 2$\farcm$5$\times$2$\farcm$5 for IGR J00234+6141 and 
10$'$$\times$10$'$ for the other objects.
In all cases, North is up and East to the left.}
\end{figure*}

\begin{table*}[t!]
\caption[]{Log of the spectroscopic observations presented in this paper.}
\begin{center}
\begin{tabular}{llccccc}
\noalign{\smallskip}
\hline
\hline
\noalign{\smallskip}
\multicolumn{1}{c}{Object} & \multicolumn{1}{c}{Date} & Telescope & 
Mid-exposure & Grism & Slit & Exposure \\
 & & & time (UT) & number & (arcsec) & time (s) \\
\noalign{\smallskip}
\hline
\noalign{\smallskip}

IGR J00234+6141 & 01 Oct 2005 & 1.5m Loiano & 18:41:21 & \#4 & 2.0 & 
2$\times$1800 \\

IGR J01583+6713 & 23 Dec 2005 & 1.5m Loiano & 17:41:25 & \#4 & 2.0 & 
2$\times$1200 \\

IGR J03532$-$6829 (=PKS 0352$-$686) & 05 Oct 1994 & 2.2m ESO/MPG & 
09:17:27 & \#1 & 1.5 & 900 \\

IGR J06074+2205 & 06 Feb 2006 & 1.5m Loiano & 19:15:18 & \#4 & 2.0 &
2$\times$1200 \\

IGR J13091+1137 (=NGC 4992) & 01 Feb 2006 & 1.5m Loiano & 03:01:56 & 
\#4 & 2.0 & 900 \\

MCG +04-48-002  & 01 Oct 2005 & 1.5m Loiano & 22:24:20 & \#4 & 2.0 & 
2$\times$1200 \\

NGC 6921        & 18 Oct 2005 & 1.5m Loiano & 23:06:47 & \#4 & 2.0 & 900 \\

\noalign{\smallskip}
\hline
\noalign{\smallskip}
\end{tabular}
\end{center}
\end{table*}

The Bologna Astronomical Observatory 1.52-metre ``G.D. Cassini'' telescope 
equipped with BFOSC was used to make spectroscopic observations of the two 
brighter objects within the {\it ROSAT} error box of X--ray source 1RXS 
J002258.3+614111, the putative optical counterparts of IGR J01583+6713
and IGR J06074+2205, and galaxies NGC 4992, MCG +04-48-002 and NGC 6921 
(see Fig. 1). The BFOSC instrument uses a 1300$\times$1340 pixel EEV CCD. 
In all 
observations, Grism \#4 and a slit width of $2''$ were used, providing a 
3500--8700 \AA~nominal spectral coverage. The use of this setup secured a 
final dispersion of 4.0~\AA/pix for all spectra. The spectra of the 
putative counterparts of 1RXS J002258.3+614111 were acquired with the slit 
rotated northwards by 25$^{\circ}$ from its original E-W position in order 
to include both objects at once. The complete log of the observations is 
reported in Table 1.

After cosmic-ray rejection, the spectra were reduced, background
subtracted and optimally extracted (Horne 1986) using IRAF\footnote{IRAF
is the Image Reduction and Analysis Facility made available to the
astronomical community by the National Optical Astronomy Observatories,
which are operated by AURA, Inc., under contract with the U.S. National
Science Foundation. It is available at {\tt http://iraf.noao.edu/}}.
Wavelength calibration was performed using He-Ar lamps acquired soon 
after each spectroscopic exposure; all spectra were then flux-calibrated 
using the spectrophotometric standard BD+25$^\circ$3941 (Stone 1977). 
Finally, and when applicable, different spectra of the same object were 
stacked together to increase the S/N ratio. The wavelength calibration 
uncertainty was $\sim$0.5~\AA~for all cases; this was checked using 
the positions of background night sky lines.

Details on the spectroscopic observation concerning PKS 0532$-$686 secured 
with the 2.2m ESO/MPG telescope plus EFOSC, and on the corresponding data 
reduction, are reported in Table 1 and in Fischer et al. (1998).

\begin{figure*}[t!]
\centering{\mbox{\psfig{file=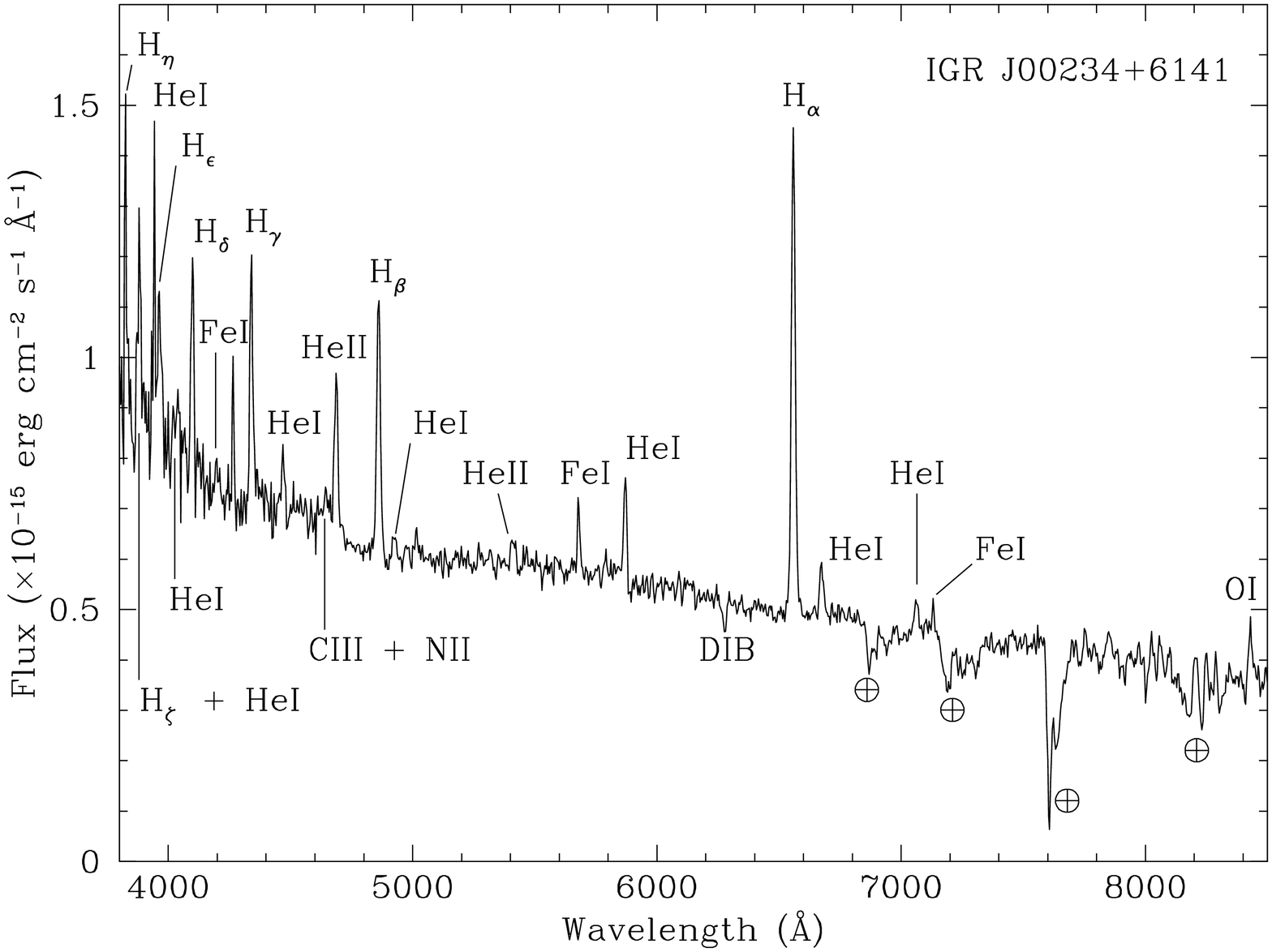,width=8.9cm,angle=270}}}
\centering{\mbox{\psfig{file=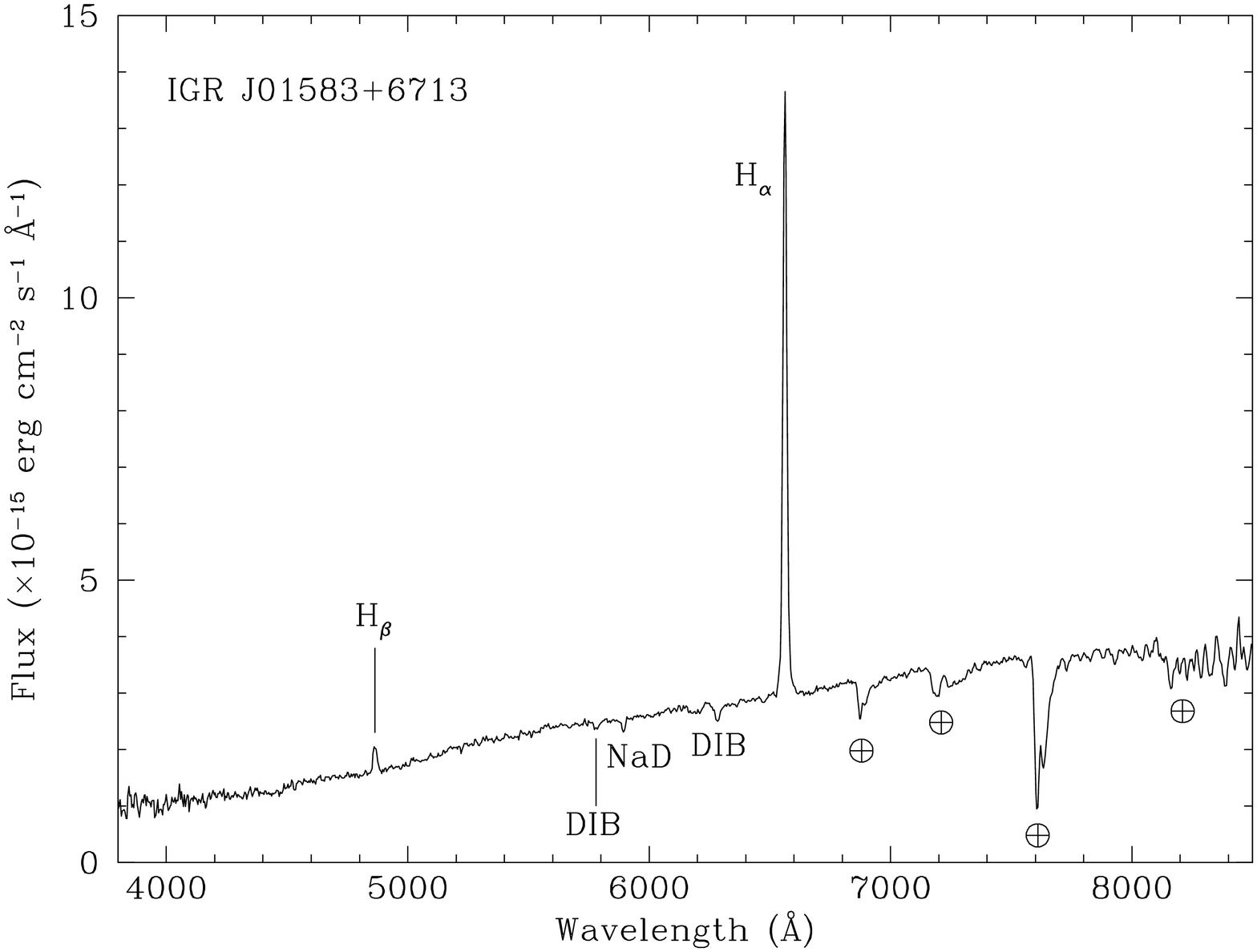,width=8.9cm,angle=270}}}
\centering{\mbox{\psfig{file=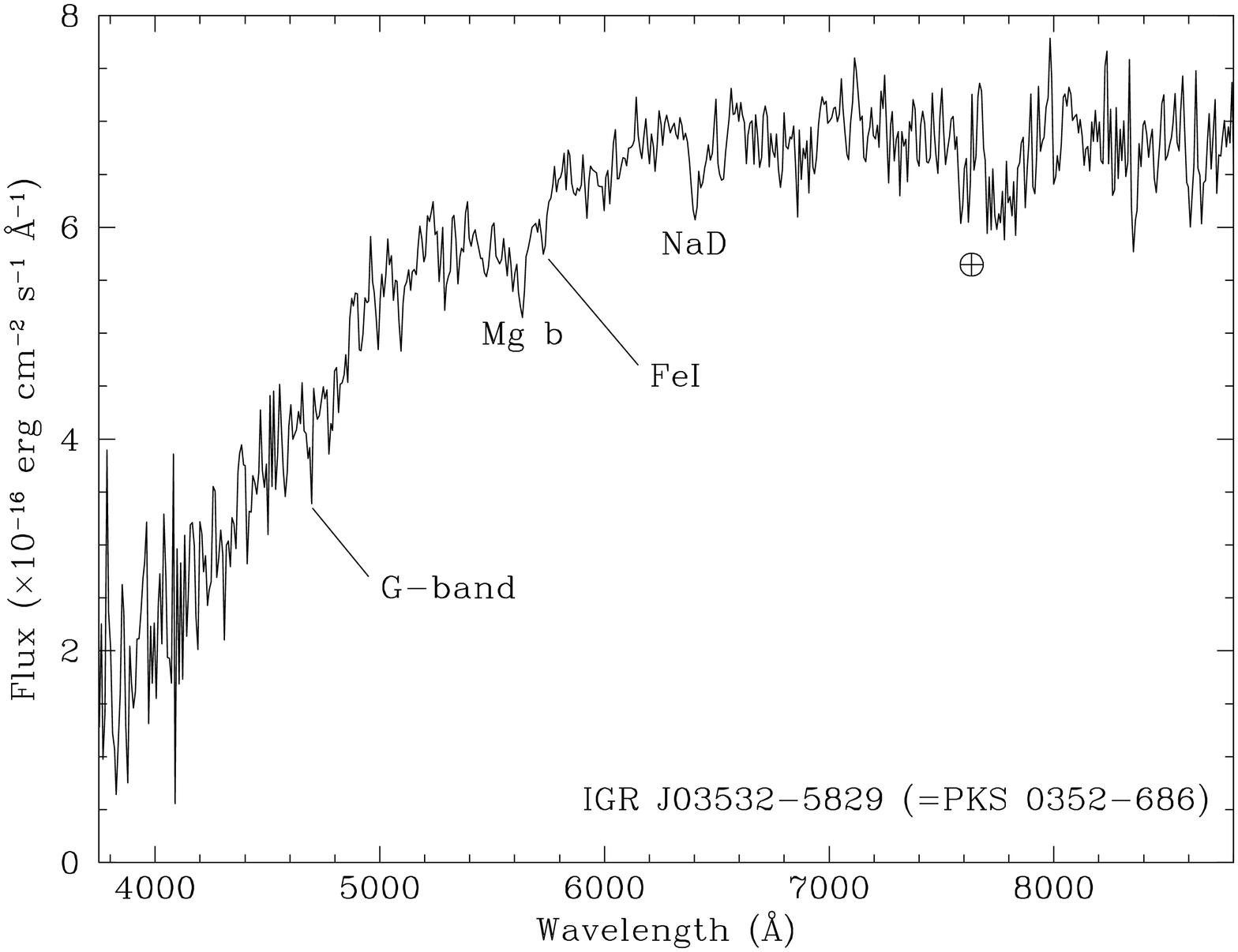,width=8.9cm,angle=270}}}
\centering{\mbox{\psfig{file=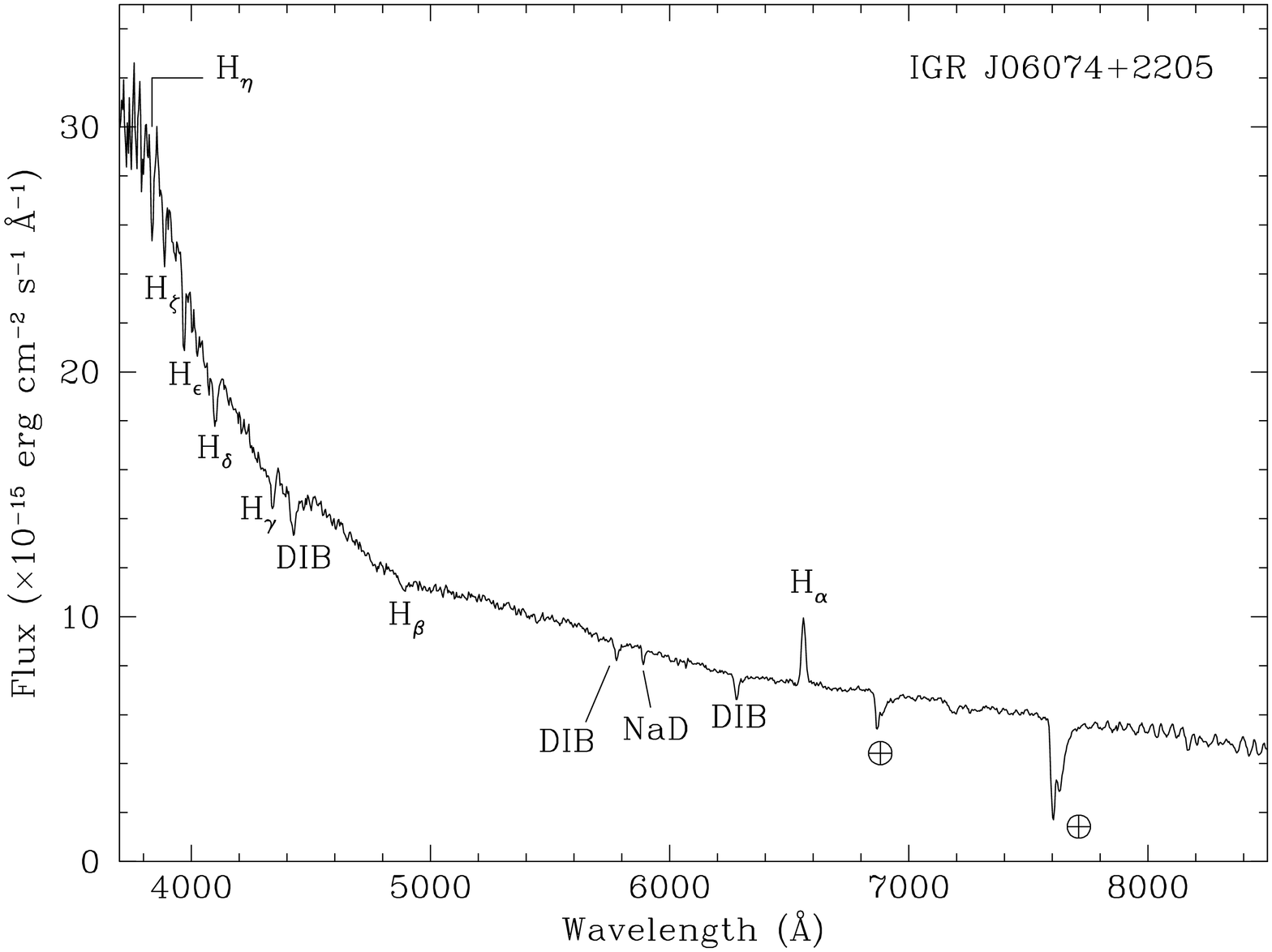,width=8.9cm,angle=270}}}
\caption{Spectra (not corrected for the intervening Galactic absorption) 
of the optical counterparts of IGR J00234+6141 (upper left panel),
IGR J01583+6713 (upper right panel), IGR J06074+2205 (lower right panel)
acquired with the Cassini telescope at Loiano, and of IGR J03532$-$6829
(=PKS 0352$-$686; lower left panel) acquired with the 2.2m ESO/MPG. For 
each spectrum the main spectral features are labeled. The symbol $\oplus$ 
indicates atmospheric telluric absorption bands.}
\end{figure*}

\begin{figure*}[t!]
\centering{\mbox{\psfig{file=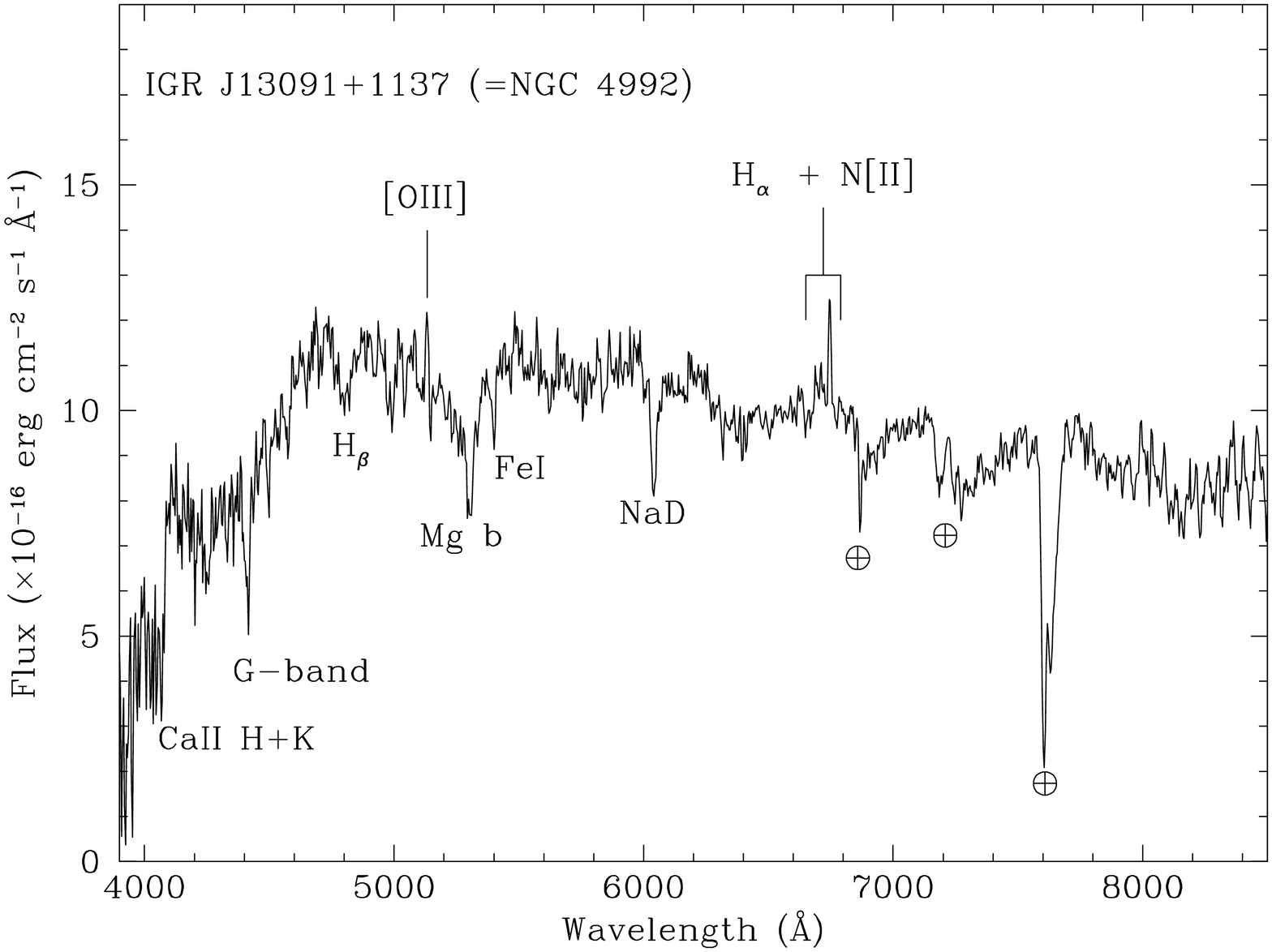,width=8.9cm,angle=270}}}
\centering{\mbox{\psfig{file=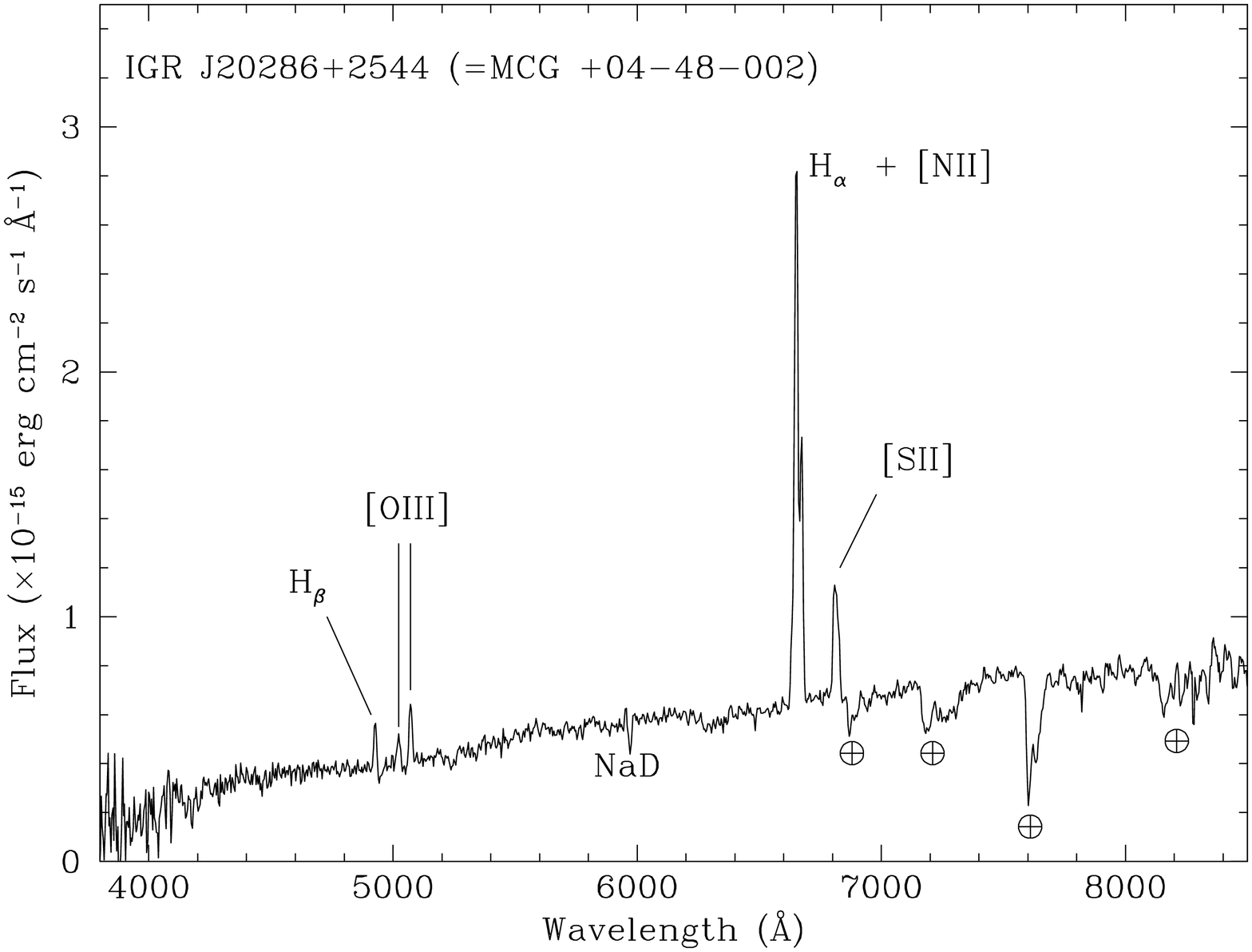,width=8.9cm,angle=270}}}
\parbox{9cm}{
\hspace{-.5cm}
\psfig{file=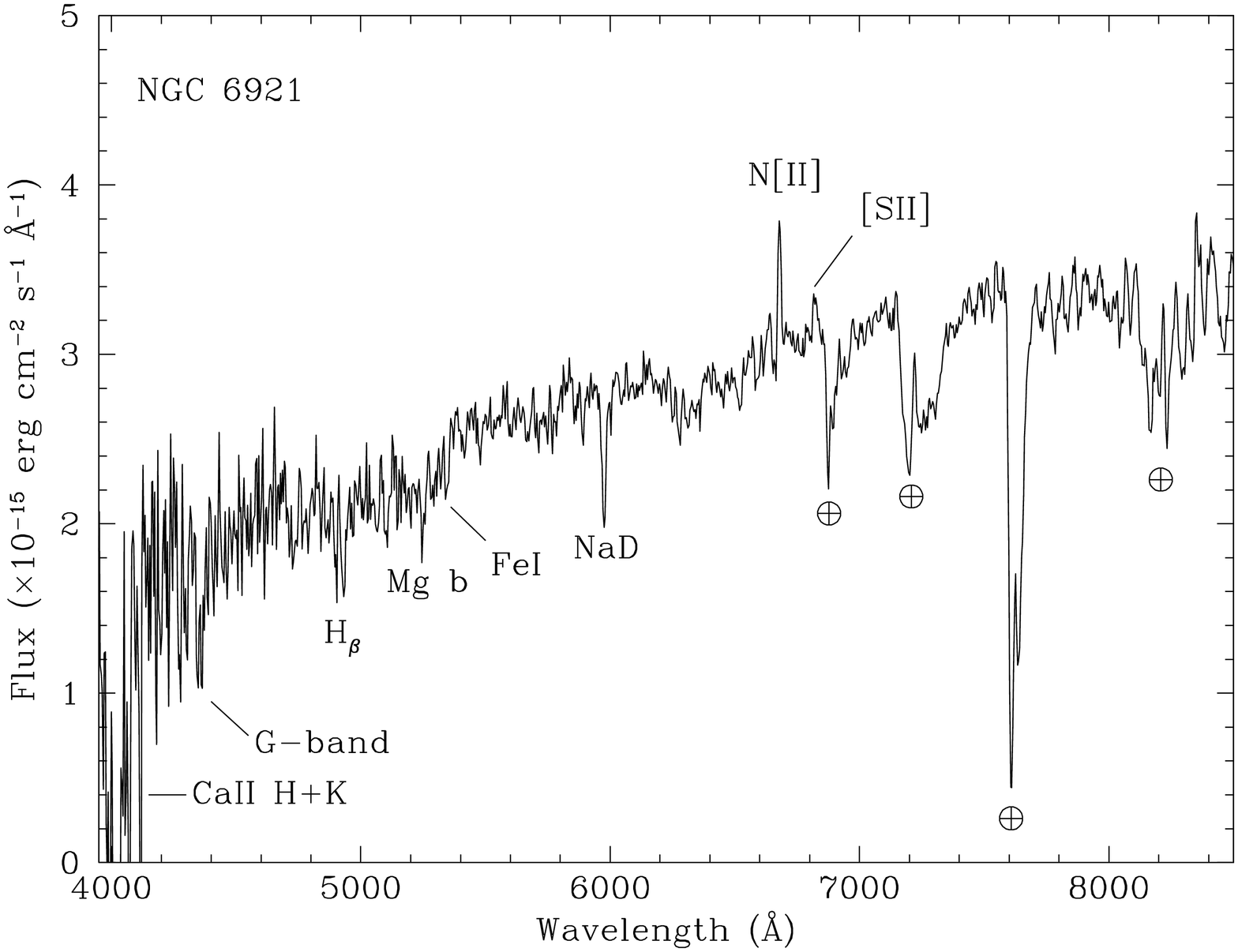,width=8.9cm,angle=270}
}
\hspace{0.5cm}
\parbox{7.5cm}{
\vspace{-1cm}
\caption{As Fig. 2, but for objects IGR J13091+1137 (=NGC 4992;
upper left panel), IGR J20286+2544 (=MCG +04-48-002; upper right panel) 
and NGC 6921 (lower left panel). All spectra were acquired 
with the Cassini telescope in Loiano.}
}
\end{figure*}

\section{Results and Discussion}

Table 2 reports the (observer's frame) wavelengths, fluxes and equivalent 
widths (EWs) of the main emission lines present in each spectrum displayed 
in Figs. 2 and 3. The line fluxes from the extragalactic sources NGC 
4992, MCG +04-48-002 and NGC 6921 were dereddened for Galactic absorption 
along their line of sight 
assuming color excesses $E(B-V)$ = 0.026 mag for the first and 
$E(B-V)$ = 0.44 mag for the other two, following the prescription of 
Schlegel et al. (1998) and assuming the Galactic extinction law of 
Cardelli et al. (1989). The spectra of these galaxies were not corrected for 
starlight contamination (see, e.g., Ho et al. 1993, 1997) given their 
limited S/N and resolution. We do not consider this to affect any of our 
conclusions. In the following we assume a cosmology with $H_{\rm 0}$ = 65 
km s$^{-1}$ Mpc$^{-1}$, $\Omega_{\Lambda}$ = 0.7 and $\Omega_{\rm m}$ = 
0.3.

\begin{table}[t!]
\caption[]{Observer's frame wavelengths, EWs (both in \AA ngstroms) and
fluxes (in units of 10$^{-15}$ erg s$^{-1}$ cm$^{-2}$) of the main emission
lines detected in the spectra of the objects reported in Figs. 2 and 3. 
For NGC 4992, MCG +04-48-002 and NGC 6921 the values are corrected for 
Galactic reddening assuming a color excess $E(B-V)$ = 0.026 mag for the 
former one and $E(B-V)$ = 0.44 mag for the latter two (from 
Schlegel et al. 1998). The error on the line positions is conservatively 
assumed to be $\pm$4 \AA, i.e., comparable with the spectral dispersion 
(see text).}
\begin{center}
\begin{tabular}{lcrr}
\noalign{\smallskip}
\hline
\hline
\noalign{\smallskip}
\multicolumn{1}{c}{Line} & $\lambda_{\rm obs}$ (\AA) & 
\multicolumn{1}{c}{EW$_{\rm obs}$ (\AA)} & \multicolumn{1}{c}{Flux} \\
\noalign{\smallskip}
\hline
\noalign{\smallskip}

\multicolumn{4}{c}{IGR J00234+6141} \\

H$_\delta$                      & 4098 &  7.0$\pm$0.7 &  5.6$\pm$0.6 \\
Fe {\sc i} $\lambda$4264        & 4264 &  3.0$\pm$0.5 &  2.2$\pm$0.3 \\
H$_\gamma$                      & 4339 &  8.6$\pm$0.6 &  6.4$\pm$0.4 \\
He {\sc i} $\lambda$4471        & 4469 &  2.1$\pm$0.4 &  1.5$\pm$0.3 \\
He {\sc ii} $\lambda$4686       & 4687 &  8.1$\pm$0.6 &  5.4$\pm$0.4 \\
H$_\beta$                       & 4861 & 13.8$\pm$0.7 &  8.7$\pm$0.4 \\
Fe {\sc i} $\lambda$5769        & 5679 &  3.0$\pm$0.3 & 1.73$\pm$0.17 \\
He {\sc i} $\lambda$5875        & 5870 &  5.5$\pm$0.4 &  3.1$\pm$0.2 \\
H$_\alpha$                      & 6559 & 37.1$\pm$1.1 & 18.4$\pm$0.6 \\
He {\sc i} $\lambda$6678        & 6674 &  4.3$\pm$0.4 &  2.1$\pm$0.2 \\
He {\sc i} $\lambda$7065        & 7063 &  1.8$\pm$0.2 & 0.82$\pm$0.12 \\
Fe {\sc i} $\lambda$7131        & 7131 &  1.1$\pm$0.2 &  0.5$\pm$0.1 \\
O {\sc i} $\lambda$8428         & 8430 &  4.1$\pm$0.8 &  1.5$\pm$0.3 \\

& & & \\

\multicolumn{4}{c}{IGR J01583+6713} \\

H$_\beta$                       & 4864 &  6.4$\pm$0.4 & 10.1$\pm$0.7 \\
H$_\alpha$                      & 6563 &   74$\pm$2   &  216$\pm$7 \\

& & & \\

\multicolumn{4}{c}{IGR J06074+2205} \\

H$_\alpha$                      & 6560 &  8.3$\pm$0.6 &   60$\pm$4 \\

& & & \\

\multicolumn{4}{c}{IGR J13091+1137} \\

$[$O {\sc iii}$]$ $\lambda$5007 & 5132 & 1.5$\pm$0.5 & 1.7$\pm$0.5 \\
$[$N {\sc ii}$]$ $\lambda$6548  & 6690 & 0.9$\pm$0.3 & 1.0$\pm$0.3 \\
H$_\alpha$                      & 6712 & 1.8$\pm$0.6 & 1.9$\pm$0.6 \\
$[$N {\sc ii}$]$ $\lambda$6583  & 6746 & 3.2$\pm$0.3 & 3.4$\pm$0.3 \\

& & & \\

\multicolumn{4}{c}{IGR J20286+2544 (=MCG +04-48-002)} \\

H$_\beta$                       & 4927 &  6.9$\pm$0.7 &  10.5$\pm$1.1 \\
$[$O {\sc iii}$]$ $\lambda$4958 & 5023 &  4.9$\pm$0.7 &   7.5$\pm$1.1 \\
$[$O {\sc iii}$]$ $\lambda$5007 & 5072 &  9.4$\pm$0.9 &  14.8$\pm$1.5 \\
$[$N {\sc ii}$]$ $\lambda$6548  & 6634 &  6.5$\pm$0.6 &   8.8$\pm$0.9 \\
H$_\alpha$                      & 6652 & 61$\pm$3     & 108$\pm$5 \\
$[$N {\sc ii}$]$ $\lambda$6583  & 6673 & 25.4$\pm$1.3 &  44$\pm$2 \\
$[$S {\sc ii}$]$ $\lambda$6716  & 6808 & 12.1$\pm$0.8 &  22.3$\pm$1.6 \\
$[$S {\sc ii}$]$ $\lambda$6731  & 6823 &  7.9$\pm$0.6 &  12.9$\pm$0.9 \\

& & & \\

\multicolumn{4}{c}{NGC 6921} \\

$[$N {\sc ii}$]$ $\lambda$6583  & 6679 & 3.2$\pm$0.3 & 27$\pm$3 \\
$[$S {\sc ii}$]$ $\lambda$6716  & 6815 & 0.35$\pm$0.05 & 2.9$\pm$0.4 \\
$[$S {\sc ii}$]$ $\lambda$6731  & 6825 & 1.18$\pm$0.18 & 9.8$\pm$1.5 \\

\noalign{\smallskip}
\hline
\noalign{\smallskip}
\end{tabular}
\end{center}
\end{table}

\subsection{IGR J00234+6141}

The westernmost of the two brightest putative counterparts of IGR 
J00234+6141 within the {\it ROSAT} error box can be discounted 
from being the counterpart of this hard X--ray source as its 
optical spectrum shows the typical absorption features of a 
normal late F-/early G-type star with no specific peculiarity at all.

The other, eastward, object shows an optical spectrum with
a host of narrow emission lines at redshift $z$ = 0 superimposed on 
a blue continuum. In particular, all of the Balmer series up to at 
least H$_\eta$, He {\sc i} and He {\sc ii}, the C {\sc iii} + 
N {\sc ii} Bowen blend, and other lines which we identified as 
O {\sc i} and 
Fe {\sc i}, are observed in emission (see Fig. 2, upper left panel). 
These spectral features are typical of a CV (e.g., Warner 1995) and 
are consistent with the independent subsequent findings of Halpern \& 
Mirabal (2006) derived from spectroscopy taken on 2006 January 24, and
the Bikmaev et al.'s (2006) analysis of their 2005 October 8-9 
spectra. 

The fact that we see the object at roughly the same optical level of the 
DSS-II-Red image suggests that it was at quiescence during our observation.
Moreover, the He {\sc ii} line detection is suggestive of a magnetic 
nature for this CV (e.g., Warner 1995), although this does not 
conclusively prove it.

Although this object lies nominally just outside of the 90\% {\it INTEGRAL} 
error circle, and it is therefore only marginally consistent with the hard 
X--ray position, its optical spectrum, together with the positional 
correlation with a bright {\it ROSAT} source (see Stephen et al. 2005, 
2006), strongly indicates that this is the optical counterpart of IGR 
J00234+6141.

Assuming an intrinsic H$_\alpha$/H$_\beta$ line ratio of 2.86
(Osterbrock 1989), the observed value ($\sim$2) indicates that 
negligible absorption is present along the line of sight to this 
source; this means that the object should be relatively 
close to Earth. Assuming an absolute magnitude M$_V \sim$ 9 and an 
intrinsic color index $(V-R)_0 \sim$ 0 mag (Warner 1995) we can give 
an estimate for the distance to IGR J00234+6141. With an optical
magnitude $R \sim$ 16.3 and no substantial further absorption along 
the line of sight, we determine that $d \sim$ 300 pc.

Using this distance estimate, we obtain 0.1--2.4 keV and 20--100 keV 
X--ray luminosities of 4.7$\times$10$^{30}$~erg s$^{-1}$ and
1.7$\times$10$^{32}$~erg s$^{-1}$ for IGR J00234+6141. These values are 
on the fainter side of the X--ray luminosity distribution for high-energy 
emitting intermediate polar (IP) CVs (see de Martino et al. 2004 and 
Suleimanov et al. 2005).

\subsection{IGR J01583+6713}

As one can see in Fig. 2, upper right panel, the optical spectrum of IGR 
J01583+6713 is characterized by an almost featureless reddened continuum 
dominated by a strong and single-peaked narrow H$_\alpha$ emission
with EW = 74$\pm$2 \AA. Further spectral characteristics are a 
weaker H$_\beta$ emission line (EW = 6.4$\pm$0.4 \AA),
along with the Na doublet at 5890 \AA~and diffuse interstellar bands in 
absorption. All features are consistent with redshift zero, indicating 
that this is a Galactic source. 
These results are comparable to those reported by 
Halpern \& Tyagi (2005a) from a spectrum acquired one day after ours,
and independently confirm the presence of absorption detected in the 
X--ray data (Kennea et al. 2005).

Confirmation of the presence of reddening comes from the 
observed H$_\alpha$/H$_\beta$ line ratio (21.4$\pm$1.6).
This, following Osterbrock (1989) implies a color excess $E(B-V)$ = 2.04.
It should be noted, however, that
if we correct the magnitudes of the optical counterpart using this
color excess, we obtain corrected color indices which are too blue
and which do not match with those of any known spectral type (Lang 1992).
We thus consider the Galactic color excess measure along the line of sight 
of IGR J01583+6713, $E(B-V)$ = 1.41 mag (Schlegel et al. 1998), a more 
acceptable estimate. This, in passing, suggests a large distance to the 
source.

We also note that the assumed reddening estimate implies, from Predehl \& 
Schmitt (1995), a column density of $\sim$1$\times$10$^{22}$ cm$^{-2}$, 
which is $\sim$10 times lower than that measured from X--ray data. 
This suggests that further absorption, produced by the material in 
accretion and local to the X--ray emitting source, is present.

The overall spectral shape strongly indicates that this is a HMXB.
However, the lack of detectable stellar absorption lines (mostly due to 
the substantial absorption along the line of sight of IGR J01583+6713)
prevents a secure spectral classification of the companion star.
Nevertheless, 
we can confidently say that a blue supergiant is ruled out due to the 
large EW of H$_\alpha$ (see e.g. the compilation of Leitherer 1988).

Next, using the optical magnitudes for this object and the Galactic 
color excess value, we find that the star has a dereddened color index 
$(B-R)_0$ = $-$0.42. This, according to Wegner (1994), means that the 
spectral type of the secondary star in the X--ray binary IGR J01583+6713 
is either O8\,III or O9\,V. If we use the absolute magnitudes and color 
indices of stars as in Lang (1992), these choices imply distances 
of 11.6 kpc and 6.4 kpc, respectively. We consider the latter as the 
more viable option because the source has Galactic longitude $l$ = 
129$\fdg$4, and a distance of 11.6 kpc from Earth would place IGR 
J01583+6713 far outside the Galaxy. Assuming a supergiant companion would 
move the system further away, thus making this hypothesis even more
unlikely.

The cases for a Low Mass X--ray Binary (LMXB) or a CV as the nature of 
this source are less 
appealing. Indeed, in the assumption that $(B-R)_0$ $\sim$ 0 $\sim$ M$_R$
for a LMXB (e.g., van Paradijs \& McClintock 1995), one finds a distance 
of 1.2 kpc for the source. An even more extreme situation is found in the 
CV case: using the same approach applied to IGR J00234+6141 (see Sect. 
4.1), we obtain a distance of 19 pc.
Neither distance can explain the large reddening observed in X--rays 
and optical and, to our knowledge, no known LMXBs or CVs show such a huge 
H$_\alpha$ emission. Besides, the comparison of the spectrum of IGR 
J01583+6713 with that of IGR J00234+6141 suggests that they belong to 
different classes of objects.

Thus, using the above distance estimate of 6.4 kpc, we find that this 
source had X--ray luminosities of 7.3$\times$10$^{34}$ erg s$^{-1}$ and 
5.2$\times$10$^{35}$ erg s$^{-1}$ in the 0.2--10 keV and 20--40 keV bands, 
respectively, during the active phase. These luminosities are comparable 
to that of the Be/X HMXB system X Per/4U 0352+309 (e.g., Haberl et al. 
1998). Therefore we conclude that IGR J01583+6713 is a transient 
Be/X HMXB.

\subsection{IGR J03532$-$6829 (=PKS 0352$-$686)}

This object does not show emission lines in its optical spectrum (see Fig. 
2, lower left panel): only a few absorption features, such as the G- and 
Mg b bands at 4304 and 5175 \AA, respectively, and the Fe {\sc i} 
$\lambda$5270 line, are seen with a redshift $z$ = 0.087$\pm$0.001, as 
reported by Fischer et al. (1998).

In order to classify the galaxy, we followed the approach of
Laurent-Muehleisen et al. (1998), finding the following relevant 
information: (1) no emission lines are detected, the upper limit to the 
EW of any emission feature being less than $\sim$5 \AA; (2) no significant 
Balmer absorption lines are detected; (3) the absorption features 
superimposed on the continuum of PKS 0352$-$686 have strengths 
generally consistent with those of other BL Lac objects (see 
Laurent-Muehleisen et al. 1998); (4) the Ca {\sc ii} break contrast at 
4000 \AA~(Br$_{\rm 4000}$), as defined by Dressler \& Shectman (1987), 
is $\sim$29\%, again similar to other BL Lac objects in the sample of 
Laurent-Muehleisen et al. (1998). 

Moreover, according to Fischer et al. (1998), the X--ray emission 
detected by {\it ROSAT} does not appear to be extended, and thus is more 
consistent with being produced by an AGN rather than by heated gas in a 
cluster. Further, substantial radio emission is detected from the galaxy.
By using the radio-to-optical 
$\alpha_{\rm ro}$ and optical-to-X--ray $\alpha_{\rm ox}$ spectral
indices as defined in Laurent-Muehleisen et al. (1999), we get 
the same indication: indeed, the values of the two parameters above
for PKS 0352$-$686 are $\sim$0.3 and $\sim$1.6, respectively, 
placing it in the domain of BL Lacs.
Taking into account the above information, we are led to classify 
PKS 0352$-$686 as a BL Lac object and as the optical counterpart 
of the hard X--ray source IGR J03532$-$6829.

The redshift of PKS 0352$-$686 implies a luminosity distance 
$d_{\rm L}$ = 428 Mpc to this galaxy. This allows us to compute the 
0.1--2.4 keV, 3--8 keV and 20--40 keV luminosities of this source
as 7$\times$10$^{43}$ erg s$^{-1}$, 1.5$\times$10$^{44}$ erg s$^{-1}$
and 1$\times$10$^{44}$ erg s$^{-1}$, respectively.
Analogously, correcting for the Galactic reddening along the PKS 
0352$-$686 line of sight by assuming a color excess $E(B-V)$ = 0.093
(Schlegel et al. 1998), we find that the absolute $B$-band magnitude
of this object is M$_B \sim$ $-$24.9.
The values above place this object in the high end of the BL Lac 
luminosity distribution (e.g., Beckmann et al. 2006).

\subsection{IGR J06074+2205}

In the optical spectrum of the putative counterpart of IGR J06074+2205
(Fig. 2, lower right panel) we confirm the presence of H$_\alpha$ in 
emission and of other Balmer lines (up to at least H$_\eta$) in 
absorption at redshift $z$ = 0. We thus support the conclusions of 
Halpern \& Tyagi (2005b) according to which this is an early-type 
emission line star and that IGR J06074+2205 is a Galactic HMXB.

The absence of He {\sc ii} lines and the weakness of He {\sc i} and light 
metal absorption lines indicates (following Jaschek \& Jaschek 1987) that 
this star is of late B spectral type. The EW of the H$_\alpha$ emission 
suggests that this is not a supergiant (see Leitherer 1988); moreover, the 
narrowness of Balmer absorption and the overall similarity with the 
optical spectrum of HD 100199 and HD 146803 (Paper III) hints that this 
star may be a giant.

Thus, assuming a spectral type B8\, III (which implies $B-R$ = $-$0.12 and 
M$_B$ = $-$1.2; Wegner 1994; Lang 1992), and using USNO-B1.0 magnitudes 
$B$ = 12.70 and $R$ = 11.29 (Monet et al. 2003), we derive a color index 
$E(B-V)$ = 1.0 and a dereddened magnitude $B_0$ = 8.9. This would place 
the star at $d \sim$ 1.0 kpc from Earth, on the near side of the Perseus 
Arm (Leitch \& Vasisht 1998).

This distance implies a 3--20 keV luminosity of 2.5$\times$10$^{34}$ erg 
s$^{-1}$ for IGR J06074+2205 during outburst, comparable to that of Be/X 
HMXB transient systems in their active phase (e.g., White et al. 1995). 
However, for the reasons illustrated in Paper III, we caution the reader 
that the association between IGR J06074+2205 and this optical 
emission-line star should still be considered as tentative, albeit likely, 
until further multiwavelength investigations (e.g., an X--ray observation
able to provide an arcsec-sized position of the high-energy source) are 
available.

\subsection{IGR J13091+1137 (=NGC 4992)}

Inspection of the optical spectrum of NGC 4992 (Fig. 3, upper 
left panel)
shows the typical features of a spiral galaxy (e.g., Kennicutt 
1998), with the Ca{\sc ii} H+K doublet, the G-band at 4304 \AA, 
$H_\beta$, the 5175 \AA~Mg~b band, the Fe {\sc i} $\lambda$5270 
line, and the 5890 \AA~Na doublet, all in absorption and
at a redshift $z$ = 0.025$\pm$0.001 (thus consistent with the
measurement of Prugniel 2006). Superimposed onto this continuum,
and at the same redshift, a few narrow (with full width at half 
maximum of $\la$600 km s$^{-1}$) emission lines are detected, the most 
prominent of which is [N {\sc ii}] $\lambda$6583 (see Table 2).

This result, combined with the observed 0.5--8 keV X--ray luminosity
(see Sazonov et al. 2005) 
suggests that NGC 4992 may belong to the class of X--ray Bright, Optically 
Normal Galaxies (XBONGs; e.g., Comastri et al. 2002). Indeed, the optical 
spectral results are reminiscent of the sample of XBONGs detected with 
{\it XMM-Newton} by Severgnini et al. (2003). 

The measured redshift corresponds to a luminosity distance
$d_L$ = 118 Mpc. This implies X--ray luminosities of 
2.0$\times$10$^{42}$ erg s$^{-1}$ and 5.7$\times$10$^{43}$ erg 
s$^{-1}$ in the 0.5--8 keV and 17--60 keV bands, respectively,
average for the luminosity range of typical AGNs 
(Beckmann et al. 2006). 

Although the optical spectrum is not discriminant about the real nature 
of this source (the observed narrow lines could be associated with
the host galaxy, which may completely outshine the AGN), the high 
X--ray luminosity and the intrinsic absorption measured by {\it Chandra},
$N_{\rm H}$ = (90$\pm$10)$\times$10$^{22}$ cm$^{-2}$ according to Sazonov 
et al. (2005), along with the {\it INTEGRAL} luminosity strongly suggest 
that the best explanation for the nature of this source is a heavily 
absorbed AGN.
In any case, to conclusively settle the identification issue, and in order 
to understand whether the emission lines observed in the optical spectrum 
are associated with the host or with the AGN activity, a deeper 
multiwavelength study is needed. This, however, is beyond the scope of 
this paper.

The nondetection of H$_\beta$ emission does not allow us to
estimate the internal reddening in the nucleus of NGC 4992.
Therefore, we are not able to apply the diagnostics of
Bassani et al. (1999) and of Panessa \& Bassani (2002)
to this case.

Under the hypothesis that the observed H$_\alpha$ emission is
produced by the host galaxy and not by the AGN, we can use the
flux of this line, corrected for Galactic absorption assuming a color 
excess $E(B-V)$ = 0.026 mag (Schlegel et al. 1998), to
estimate the star formation rate (SFR) in NGC 4992. From Kennicutt 
(1998), we determine a SFR of 0.025$\pm$0.008 $M_\odot$ yr$^{-1}$ from the 
reddening-corrected H$_\alpha$ luminosity of 
(3.2$\pm$1.0)$\times$10$^{39}$ erg s$^{-1}$. This estimate should be 
considered a lower limit to the SFR as no correction for the reddening 
local to NGC 4992 was accounted for.

In conclusion, to the best of our knowledge, NGC 4992 is the first 
confirmed XBONG detected in hard X--rays with {\it INTEGRAL}, and the 
closest XBONG detected so far. Thus, it appears as an ideal laboratory for 
the study of this enigmatic class of galaxies.

\subsection{IGR J20286+2544 (=MCG +04-48-002) and NGC 6921}

The spectrum of the galaxy MCG +04-48-002 (Fig. 3, upper right panel) 
shows a number of narrow (again, with full width at half maximum of 
$\la$600 km s$^{-1}$) emission features that can be readily identified 
with redshifted optical nebular lines. These include H$_\beta$, 
[O~{\sc iii}] $\lambda\lambda$4958,5007, H$_\alpha$, [N~{\sc ii}]
$\lambda\lambda$6548,6583, and [S~{\sc ii}] $\lambda\lambda$6716,6731.
All identified emission lines yield a redshift of $z$ = 0.013$\pm$0.001, 
in agreement with Paturel et al. (2003). The NaD doublet in 
absorption is also detected at the same redshift.

Concerning the spectrum of NGC 6921 (Fig. 3, lower left panel), we find it 
to be a typical spectrum of a normal spiral galaxy, with Ca H+K, G-band, 
H$_\beta$, Mg b, Fe {\sc i} $\lambda$5270 and NaD in absorption, plus [N 
{\sc ii}] $\lambda$6583 and [S {\sc ii}] $\lambda\lambda$6716,6731 in 
emission. All of these features are at a redshift of $z$ = 0.014$\pm$0.001, 
fully consistent with that of Paturel et al. (2003). Given the 
substantially lower X--ray flux (as measured with XRT) from the nucleus 
of NGC 6921 with respect to that from MCG +04-48-002, we conclude that 
the actual optical counterpart of IGR J20286+2544 is the galaxy MCG
+04-48-002, although we cannot exclude a marginal contribution from
NGC 6921 to the total hard X--ray emission detected with ISGRI.

The diagnostics of Ho et al. (1993, 1997) seem to suggest 
that MCG +04-48-002 is a Starburst/H {\sc ii} galaxy: indeed, 
the line ratios [N~{\sc ii}]/H$_\alpha$, [S~{\sc ii}]/H$_\alpha$,
and [O~{\sc iii}]/H$_\beta$, together with the nondetection of 
[O~{\sc i}] $\lambda$6300 in emission (the 3$\sigma$ upper limit to
the flux of this line is 7$\times$10$^{-16}$ erg cm$^{-2}$ s$^{-1}$),
indicates that this galaxy falls in the Starburst/H {\sc ii} locus.

In spite of this, 
using the cosmology described above and the more accurate redshift
of Paturel et al. (2003), we find that the luminosity distance
to the galaxy MCG +04-48-002 is $d_L$ = 66.2 Mpc, and that its 
20--100 keV X--ray luminosity is 2.1$\times$10$^{43}$~erg s$^{-1}$.
This value is very high for a Starburst (David et al. 1992),
whereas it would be around average for a 
Type 2 Seyfert galaxy (see, e.g., Beckmann et al. 2006).
The measured value for the X--ray luminosity of MCG +04-48-002 
is thus comparable with that of ``classical'' AGNs.
Further, the detection of radio and hard X--ray emission and the 
{\it ROSAT} 0.1--2.4 keV nondetection suggest that this object is
an obscured AGN.

This conclusion is further supported by the X--ray/FIR and
[O {\sc iii}]/FIR flux ratios for this object: following
the approach of Panessa \& Bassani (2002), we find that its
IRAS 25$\mu$m+60 $\mu$m flux, F$_{\rm FIR}$ = 6.9$\times$10$^{-10}$~erg 
cm$^{-2}$ s$^{-1}$ (Sanders et al. 2003), its 2--10 keV flux 
(2.3$\times$10$^{-10}$~erg 
cm$^{-2}$ s$^{-1}$; Landi et al., in preparation) and its unabsorbed 
[O {\sc iii}] $\lambda$5007 line flux (9.4$\times$10$^{-13}$~erg 
cm$^{-2}$ s$^{-1}$; see below for the correction for internal absorption), 
place the nucleus of MCG +04-48-002 in the loci populated by Seyfert 2, 
rather than Starburst, galaxies (Fig. 2 and 3 of Panessa \& Bassani 
2002). In addition, the X--ray/[O~{\sc iii}]$_{\rm 5007}$ ratio, 
$\sim$2.5, indicates that this source is in the Compton-thick regime 
(see Bassani et al. 1999).

The strength of the optical emission lines of MCG +04-48-002, after 
accounting for Galactic and intrinsic absorptions, can be used to 
estimate the SFR and metallicity of this galaxy. First, a
correction for Galactic reddening has been applied (we assumed a color 
excess $E(B-V)$ = 0.44 mag following Schlegel et al. 1998). Next, 
considering an intrinsic Balmer decrement of H$_\alpha$/H$_\beta$ = 2.86 
(Osterbrock 1989) and the extinction law of Cardelli et al. (1989), the 
observed flux ratio H$_\alpha$/H$_\beta$ = 10.3 implies an internal color 
excess $E(B-V)$ = 1.30 mag (in the galaxy rest frame). Following Kennicutt 
(1998), we determine a SFR of 9.3$\pm$0.9 $M_\odot$ yr$^{-1}$ from the
reddening-corrected H$_\alpha$ luminosity of 
(1.18$\pm$0.06)$\times$10$^{42}$ erg s$^{-1}$. 

Although the [O {\sc ii}] emission line, if present, falls outside the 
actual wavelength range covered by our Loiano spectra of MCG +04-48-002, 
the detection of [O {\sc iii}] and H$_\beta$ allows us to infer a range for 
the gaseous oxygen abundance in this galaxy. Following Kobulnicky et al. 
(1999), the $R_{\rm 23}$ parameter, defined as the ratio between [O {\sc 
ii}] + [O {\sc iii}] and H$_\beta$ line fluxes, gives 7.0 $<$ 12 + log 
(O/H) $<$ 9.0. Considering the intrinsic luminosity of the source (it has 
rest-frame absolute $B$-band magnitude M$_B$ = $-$21.36 mag; Prugniel 
2005) and its [O {\sc iii}]/[N~{\sc ii}] ratio ($\sim$1.6), all of this 
information points to the fact that this galaxy is on the high-metallicity 
side of the range reported above, i.e., 12 + log (O/H) $>$ 8.3.
We caution the reader that the [O {\sc iii}] and [N {\sc ii}]
lines can be contamined by AGN activity, so this may impact on 
the metalliticy lower limit derived above.

In conclusion, we put forward the hypothesis that this 
Starburst/H {\sc ii} galaxy, similarly to NGC 4945 (L\'{i}pari 
et al. 1997), hides an obscured Seyfert 2 nucleus, which can be 
seen only at hard X--rays and infrared/radio wavelengths.
If this is true, this galaxy represents a further example of an 
optically elusive AGN, similar to NGC 4992 (see Sect. 4.5).

\section{Conclusion}

In our continuing work aimed at the identification of unknown 
{\it INTEGRAL} sources by means of optical spectroscopy (Papers 
I-III), we have identified and studied six more hard X--ray 
objects by using the Astronomical Observatory of Bologna in Loiano 
(Italy) and by reexamining spectroscopic data acquired at the 2.2m 
ESO/MPG telescope in La Silla (Chile).

We determined the nature of the sources as follows: (i) IGR J00234+6141 
is a CV, probably an IP, caught during a quiescent phase and located at 
$d \sim$ 300 pc from Earth; (ii) IGR J01583+6713 is a transient Be/X 
HMXB located at $\sim$6.4 kpc from Earth; (iii) IGR 
J03532$-$6829 (=PKS 0352$-$686) is a BL Lac at redshift $z$ = 0.087;  
(iv) IGR 06074+2205 is likely a Be/X HMXB located at $\sim$1 kpc from 
Earth, although soft X--ray observations should be made to provide a
conclusive test on this identification;
(v) IGR J13091+1137 is an XBONG probably hiding an absorbed AGN,
at redshift $z$ = 0.025 and with SFR $\sim$ 0.03 $M_\odot$ yr$^{-1}$; 
(vi) IGR J20286+2544 (=MCG +04-48-002) is a Starburst/H {\sc ii} galaxy 
located at $z$ = 0.013, with nearly solar metallicity and SFR $\sim$ 9 
$M_\odot$ yr$^{-1}$, hiding an obscured Type 2 Seyfert nucleus in the 
Compton-thick regime.

These results once more stress the importance of {\it INTEGRAL} for the 
detection and the study of hard X--ray emission not only from X--ray 
binaries, but also from CVs and extragalactic objects. Moreover, the 
findings presented here indicate that {\it INTEGRAL} is revealing a 
population of `buried' AGNs, the activity of which is hidden in the 
optical by local absorption and/or by the luminosity of their host 
galaxies.

\begin{acknowledgements}

We thank the anonymous referee for useful remarks which helped us to 
improve the paper. We also thank Stefano Bernabei, Ivan Bruni, Antonio 
De Blasi and Roberto Gualandi for the assistance at the telescope in 
Loiano and for having performed part of the observations in Loiano 
as a Service Mode run, and Jules Halpern and Retha Pretorius for 
comments and discussions.
This research has made use of the NASA Astrophysics 
Data System Abstract Service, of the NASA/IPAC Extragalactic Database 
(NED), and of the NASA/IPAC Infrared Science Archive, which are operated 
by the Jet Propulsion Laboratory, California Institute of Technology, 
under contract with the National Aeronautics and Space Administration. 
This research has also made use of the SIMBAD database operated at CDS, 
Strasbourg, France, and of the HyperLeda catalogue operated at the 
Observatoire de Lyon, France.
The authors acknowledge the ASI financial support via grant I/R/046/04.

\end{acknowledgements}

\end{document}